\documentclass[aps,floats,twocolumn,epsf,prx,superscriptaddress,showpacs,longbibliography,floatfix]{revtex4-2}

\usepackage[utf8]{inputenc}
\usepackage{graphicx}
\usepackage{epstopdf}
\usepackage{amsmath}
\usepackage{booktabs}
\usepackage[colorlinks=true,urlcolor=blue,citecolor=blue,linkcolor=blue,breaklinks=true]{hyperref}
\usepackage{color}
\usepackage{bm}
\usepackage{dsfont}
\usepackage[dvipsnames]{xcolor}
\usepackage{multirow}
\usepackage{amsfonts}
\usepackage[normalem]{ulem}
\usepackage{bm}
\usepackage{float}
\usepackage{placeins}

\usepackage{orcidlink} 

\newcommand{\braket}[1]{\langle #1 \rangle}

\definecolor{myRed}{RGB}{150, 0, 24}

\newcommand{\Ce}{CeRu$_4$Sn$_6$}
\newcommand{\CBP}{Ce$_3$Bi$_4$Pd$_3$}

\begin{document}

\preprint{PRB/123-QED}

\title{Weyl nodes in \Ce\ studied by dynamical mean-field theory}

\author{Jor\=unas Dobilas\,\orcidlink{0000-0002-8864-100X}}
\email{jorunas.dobilas@ftmc.lt}
\affiliation{Department of Functional Materials and Electronics, FTMC, 10257 Vilnius, Lithuania}
\affiliation{Institute of Solid State Physics, TU Wien, 1040 Vienna, Austria}

\author{Martin Bra\ss\orcidlink{0000-0002-4347-6987}}
\affiliation{Institute of Solid State Physics, TU Wien, 1040 Vienna, Austria}

\author{Frank T. Ebel\orcidlink{0009-0005-0479-0243}}
\affiliation{Institute of Solid State Physics, TU Wien, 1040 Vienna, Austria}

\author{Silke Paschen\orcidlink{0000-0002-3796-0713}}
\affiliation{Institute of Solid State Physics, TU Wien, 1040 Vienna, Austria}

\author{Karsten Held\orcidlink{0000-0001-5984-8549}}
\email{held@ifp.tuwien.ac.at}
\affiliation{Institute of Solid State Physics, TU Wien, 1040 Vienna, Austria}

\date{\today}

\begin{abstract}
The heavy fermion compound CeRu$_4$Sn$_6$ was recently shown to exhibit a spontaneous nonlinear Hall effect, indicating its topological nature. This is consistent with the lack of inversion symmetry that allows for the existence of Weyl nodes. Here, we employ density functional theory combined with dynamical mean-field theory, which is state-of-the-art for strongly correlated materials, and study the topology of CeRu$_4$Sn$_6$. We find five inequivalent Weyl nodes of either type I or II, each having either eight or sixteen symmetry-related replicas. These Weyl nodes bridge the Kondo insulating gap, which is a direct but not an indirect gap.
The Weyl points closest to the Fermi level are situated only 0.5\,meV below it, and have a very flat dispersion.
Our \textit{ab initio} results establish \Ce{} as a model system for investigating the interplay between strong electronic correlations and nontrivial topology. These findings provide a theoretical foundation for future studies of quantum transport and interaction-driven topological phases in heavy-fermion systems.
\end{abstract}

\maketitle
 
\section{\label{sec:Introduction}Introduction}

Quantum materials \cite{Kei17.1} with nontrivial electronic topology \cite{Wit14.1,Yan17.2,Arm18.1,Ver19.1} are in high demand for next-generation electronic and quantum devices \cite{Liu19.1,Jin23.1}.
Strongly correlated electron systems represent a fertile ground for emergent quantum phases, including heavy fermion metals, Kondo insulators, and unconventional superconductors \cite{paschen2021quantumRev}. When strong electronic interactions interplay with crystalline symmetries, correlation-driven topological phases may arise \cite{Che24.1}. However, the many-body physics that leads to these interesting quantum phenomena also makes them challenging to describe theoretically, in particular using {\it ab initio} methods without any adjustable parameters.

A recently established strongly correlated topological phase is the Weyl-Kondo semimetal \cite{Dzs17.1,Lai18.1,Gre20.1,Dzs21.1}, with \CBP\ representing its first experimental realization \cite{Dzs17.1,Dzs21.1}.
Theoretically, the Weyl-Kondo semimetal phase was introduced through a study of the periodic Anderson model on a noncentrosymmetric lattice. The space group symmetry constraints on the Kondo-driven low-energy electronic states create Weyl nodes of heavy quasiparticles in the immediate vicinity of the Fermi level \cite{Lai18.1,Gre20.1}.

Here, we study the compound CeRu$_4$Sn$_6$, which crystallizes in a noncentrosymmetric tetragonal crystal structure of the space group $I\bar{4}2m$ (No.\ 121) \cite{ZumdickPoettgen1999}. Early electrical resistivity \cite{Win12.1} and optical conductivity data \cite{Gur13.1} on single crystals, and NMR data on powder \cite{Bru06.1} led to the interpretation of \Ce\ as a ``failed'' Kondo insulator. With decreasing temperature, a Kondo insulator gap starts to form but ultimately fails to develop fully, leading to a semimetallic ground state. Later, from a treatment using density functional theory (DFT) plus the Gutzwiller approximation, it was proposed that \Ce\ might be a correlated Weyl semimetal \cite{XuHeavyFermions}. Most recently, the hallmark of a strongly correlated Weyl semimetal---a giant spontaneous nonlinear Hall effect \cite{Dzs21.1,Kir24.1} originating from the Berry curvature monopoles at Weyl nodes near the Fermi energy \cite{Dzs17.1,Lai18.1,Gre20.1}---was indeed identified in low-temperature experiments on \Ce\ \cite{Kir24.2x}. 

Furthermore, the low-temperature electronic specific heat of \Ce\ was understood as a convolution of a quantum critical and a $T^3$ contribution \cite{Kir24.2x}. The latter is another key signature of the Weyl-Kondo semimetal \cite{Dzs17.1,Lai18.1,Gre20.1}: It is direct evidence of the linear Weyl dispersion, with Weyl nodes pinned to the Fermi energy,
and a prefactor that allows for the extraction of the Weyl velocity. Just as in \CBP\ \cite{Dzs17.1}, the Weyl velocity of \Ce\ is experimentally found to be less than 1\,km/s, orders of magnitude lower than in noninteracting systems \cite{Kir24.2x}.

Motivated by these findings, we  investigate the situation employing state-of-the-art DFT plus dynamical mean field theory (DMFT) \cite{Metzner1989,Georges1992,Jarrell1992,Georges1996,Juelich2014,held2007electronic}, and follow the Berry curvature 
to detect the Weyl nodes.
We find that the hybridization between Ce 4$f$ and conduction electrons leads to a direct (but not indirect) gap throughout the Brillouin zone. However, this direct gap is bridged along some low symmetry directions by $64$ Weyl nodes.

The paper is organized as follows: In Section~\ref{Sec:CompDetails} we briefly recapitulate the computational methods used, i.e., the DFT+DMFT approach in Section~\ref{Sec:DFT+DMFT} and the Weyl node search algorithm in Section~\ref{Sec:WeylSearch}. In Section~\ref{Sec:Results} we present the results obtained starting with the DFT+DMFT spectra and self energies in Section~\ref{Sec:resultsDFTDMFT} before turning to the topological properties in Section~\ref{Sec:ResultsTopology}. Section~\ref{Sec:Conclusion} summarizes our results; and Appendix~\ref{App:D2d} lists the symmetry operations for obtaining all $64$ Weyl nodes from the five inequivalent ones.

\begin{figure*}
    \centering
    \includegraphics[width=0.9\linewidth]{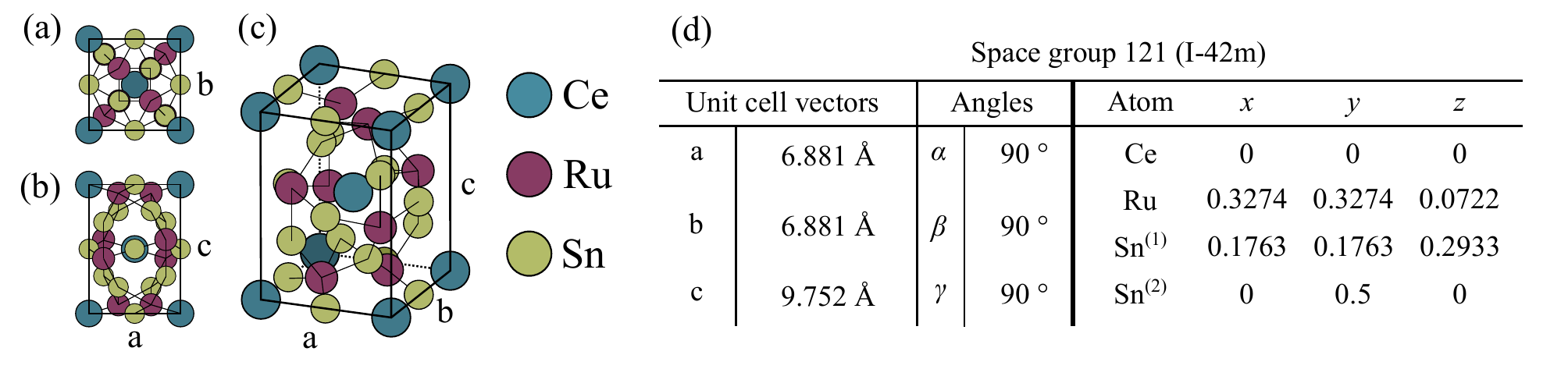}
    \caption{Crystal structure of \Ce. Panels (a) and (b) show projections of the crystal structure along the (001) and (010) directions, respectively, (c) a three-dimensional view of the unit cell. In Table (d), the atomic coordinates are expressed in units of the lattice vectors. The crystallographic parameters are taken from \cite{ZumdickPoettgen1999}.}
    \label{CeRu4Sn6Crystal}
\end{figure*}

\section{Computational details}
\label{Sec:CompDetails}
\subsection{DFT+DMFT}
\label{Sec:DFT+DMFT}
As a first step of our DFT+DMFT study, the electronic band structure of \Ce\ is calculated using the (FPLO) code \cite{koepernik1999} and, for comparison, the \textsc{WIEN2k} \cite{Wien2K} full potential linearized augmented plane wave (FP-LAPW) package. Details of the crystal structure of \Ce\ \cite{ZumdickPoettgen1999} are summarized in Fig.~\ref{CeRu4Sn6Crystal}.
Following previous studies \cite{Wissgott,XuHeavyFermions}, we calculate the electronic band structure using the local density approximation (LDA) with the Perdew-Wang exchange-correlation potential~\cite{perdew1992} and a dense (6 × 6 × 6) $\bm{k}$ mesh to determine the electronic band structure. After the calculation converges, we take the resulting ground state density as the starting point for another self-consistent computation, where we increase the $\bm{k}$ mesh to (12 × 12 × 12). This computation converges immediately, which confirms that the initial $\bm{k}$ mesh is sufficient. All calculations are fully relativistic, especially including spin-orbit coupling (SOC).

To simplify our DMFT calculations and to improve numerical accuracy, we employ a tight-binding Hamiltonian based on the FPLO calculation. The tight-binding parameterization is motivated by the fact that subsequent calculations involve extensive sampling at different $\bm{k}$ points, where the reduced basis significantly improves computational efficiency. The new basis consists of $90$ orbitals, giving a $90\times90$ Hamiltonian matrix. These orbitals are
\begin{itemize}
    \item Ce $4f$ orbitals (total of $14$: $6$ $f_{5/2}$ and $8$ $f_{7/2}$ orbitals),
    \item Ru $4d$ orbitals (total of $4\cdot10=40$: $4\cdot6$ $d_{5/2}$ and $4\cdot4$ $d_{3/2}$ orbitals),
    \item Sn $5p$ orbitals (total of $6\cdot6=36$ orbitals: $6\cdot4$ $p_{3/2}$ and $6\cdot2$ $p_{1/2}$ orbitals).
\end{itemize}
The energy range spanned by all orbitals extends from $-5\,\mathrm{eV}$ to $5\,\mathrm{eV}$, which is sufficient to capture the relevant electronic behavior near the Fermi level.

This  tight-binding Hamiltonian is supplemented by a Coulomb interaction $U$ for the Ce $4f_{5/2}$ manifold. Since the Ce $4f_{7/2}$ bands are above the Fermi energy, we treat their interaction in a static way as a Hartree-like shift and only that of the  Ce $4f_5/2$ manifold dynamically. The intraorbital ($U$) and interorbital ($V$) Coulomb interaction potentials are set to the same value of $5.5\,\mathrm{eV}$ for the Ce $4f_{5/2}$ bands, consistent with values used in the literature \cite{Wissgott}. Additionally, the Hund's coupling is set to zero, which is justified to some extent by the fact that Ce is in a $4f^1$ configuration. Hence, the Hund's exchange is only relevant for excited states and multiplet admixing. To avoid double counting correlation effects that are partially included in the DFT functional, these are subtracted using the Anisimov formula $V_{\mathrm{DC}}=U (\langle n \rangle -\frac{1}{2})$ \cite{anisimov1991band}. For \Ce, with a Ce $4f$ shell filling of $\langle n \rangle\approx 0.73$ from DFT calculations, this correction is $V_{\mathrm{DC}}=1.27\,\mathrm{eV}$.

\begin{figure}
    \centering
    \includegraphics[width=1\linewidth]{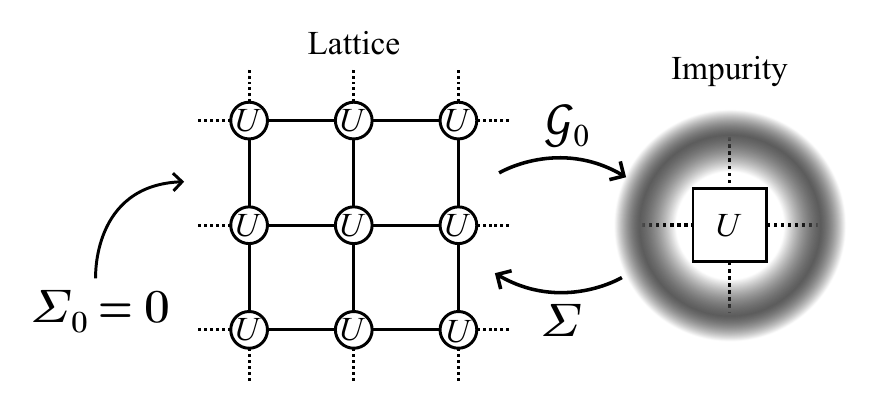}
    \caption{Self-consistency cycle of dynamical mean-field theory. The lattice model (left) is computed using the initial self-energy $\mathit{\Sigma}_0=0$ with on-site Coulomb potential $U$. The dynamic Weiss field ($\mathcal{G}_0$) is then extracted from the lattice model and used as input in the Anderson impurity model (right), from which the local self-energy $\mathit{\Sigma}$ is obtained. This self-energy is then used in turn as input in the lattice model to update $\mathcal{G}_0$. The cycle is repeated until the convergence of the self-energy is achieved.}
    \label{DMFT Cycles}
\end{figure}

Subsequently the previously defined Hamiltonian is solved by DMFT using the open-source code \textsc{w2dynamics}~\cite{Wallerberger2019}, with continuous-time quantum Monte Carlo (QMC) in the hybridization expansion~\cite{Gull2011} as a DMFT impurity solver.
In DMFT, the complex many-body problem is mapped onto an effective Anderson impurity model that is solved self-consistently \cite{Georges1992,Jarrell1992}, see the schematics Fig.~\ref{DMFT Cycles}. This captures the local electronic correlations of the Ce 4$f$ electrons while preserving the initial character of the conduction electrons. The combined DFT+DMFT \cite{Anisimov1997,Lichtenstein1998,held2007electronic} approach enables the extraction of momentum-resolved spectral functions, self-energies, and renormalized bands for real materials \cite{Juelich2014,held2007electronic}. Furthermore, it is possible to extract topological features from the renormalized band structures, such as Weyl points or surface states for finite-volume systems.

The DMFT loop is iterated for $10$ self-consistency steps. In each step, $2\times10^8$ QMC warm-up and $10^7$ measurement sweeps are performed. This procedure is independently repeated at multiple temperatures $T$ to capture the thermal evolution of the correlated electronic structure. The calculations are in the paramagnetic phase, which correctly captures the experimental situation \cite{Kir24.2x}. After completion of the DMFT iterations, we obtain Green's functions $G(\mathrm{i}\omega_n)$ and self-energies $\mathit{\Sigma}(\mathrm{i}\omega_n)$ at Matsubara frequencies $\mathrm{i}\omega_n=(2n+1)\pi/\beta$ for the correlated Ce $4f_{5/2}$ orbitals. These are then analytically continued to real frequencies $\omega$ using the open-source maximum entropy analytical continuation package \textsc{ana\_cont}~\cite{KAUFMANN2023108519}. The spectral function $A(\omega)$ is connected to $G(\omega)$ through the relation $A(\omega)=-\frac{1}{\pi}\mathrm{Im}G(\omega)$. 

To represent the many-body effects of DMFT, we further employ a quasiparticle renormalization of the DFT bands by expanding Green's functions around the Fermi level. The quasiparticle weight $Z$ for the different Ce $4f_{5/2}$ states is calculated from the real part of the self-energy:
\begin{equation}
    Z=\left(1-\left.\frac{\partial\mathrm{Re}\mathit{\Sigma}(\omega)}{\partial\omega}\right|_{\omega=0}  \right)^{-1}\; .
\end{equation}

The quasiparticle representation then yields the renormalized Hamiltonian matrix
\begin{equation}
     H_{\mathrm{qp}}(\bm{k})=\sqrt{Z} \left(H(\bm{k}) + \mathrm{Re}\mathit{\Sigma}(0) -\mu \mathbb{I} -V_{\mathrm{DC}}\right)\sqrt{Z}.
    \label{renormalized}
\end{equation}
Here $H(\bm{k})$ is the original tight-binding Hamiltonian matrix; $\mathbb{I}$ the identity matrix; $Z$, $\mathrm{Re}\mathit{\Sigma}(0)$ and $V_{\mathrm{DC}}$ are in (diagonal) matrix form and have non-zero elements only for the Ce $4f_{5/2}$ manifold. Using $H_{\mathrm{qp}}(\bm{k})$, the quasiparticle Green's function
\begin{equation}
    G_{\textrm{qp}}(\omega,\bm{k})=Z\cdot[\omega-H_{\mathrm{qp}}(\bm{k})]^{-1}
\end{equation}
is then used to compute $\bm{k}$-resolved quasiparticle spectral functions, and to determine the topology of the DFT+DMFT electronic structure.

\subsection{Weyl node search algorithm}
\label{Sec:WeylSearch}
An important quantity defining the Weyl nodes is the Berry curvature $\bm{\mathit{\Omega}}_n(\bm{k})$, which characterizes the geometrical structure of Bloch bands. For a given band $n$, the Berry curvature is defined as a curl of the Berry connection $\bm{A}_n(\bm{k})$ \cite{Hassani2017}:

\begin{equation}
    \bm{\mathit{\Omega}}_n(\bm{k})=\nabla \times \bm{A}_n(\bm{k}),
\end{equation}
with

\begin{equation}
    A_n(\bm{k})=\mathrm{i} \langle u_{n\bm{k}}|\nabla_{\bm{k}}|  u_{n\bm{k}}\rangle.
\end{equation}

Here $u_{n\bm{k}}$ is the periodic part of the Bloch function for the $n$-th band. The topological ``charge" of the Weyl nodes is defined by the Chern number, which is computed by integrating $\mathit{\Omega}_n(\bm{k})$ over a closed surface $S$:
\begin{equation}
    \mathcal{C}_n=\frac{1}{2\mathrm{\pi}}\oint_S \bm{\mathit{\Omega}}_n(\bm{k}) \mathrm{d}\bm{S}.
\end{equation}
Weyl nodes come in pairs, characterized by their Chern numbers $\mathcal{C}_n$: $+1$ for a source and $-1$ for a sink of the Berry curvature $\bm{\mathit{\Omega}}_n(\bm{k})$ in $\bm{k}$ space. This behavior is analogous to magnetic monopoles or electric charges in real space. 

There are various methods to find Weyl nodes. The brute-force approach is to take the full lattice cell, subdivide it into smaller pieces, and compute the Berry flux in each cell. If they are small enough to have only one Weyl node per cell, it is possible to pin-point the location of these nodes by continuously making the cells smaller. However, this method is numerically demanding, especially if the nodes are relatively close to each other; and very small cells can be prone to numerical inaccuracies when calculating the Berry flux.
\begin{figure}[tb]
    \centering
    \includegraphics[width=0.8\linewidth]{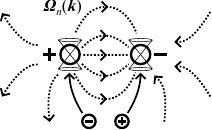}
    \caption{Schematic representation of the Weyl node search algorithm~\cite{MartinWeyl}. Randomly placed ``particles" with Chern positive or negative ``charges" ($\oplus$, $\ominus$) are created and follow the Berry curvature ``field lines" $\bm{\mathit{\Omega}}_n(\bm{k})$ (dotted lines) toward the Weyl points (symbolized by a circle with a cone dispersion). Depending on the ``charge" of the ``particles", they will move towards Weyl nodes with Chern numbers $+1$ or $-1$, respectively.}
    \label{NodeAlg}
\end{figure}

We use a more efficient method that is visualized in Fig.~\ref{NodeAlg} and is available as part of the \textsc{TightBindingToolBox} \cite{MartinWeyl}. It has been employed successfully to identify Weyl points in the heavy fermion compound Ce$_3$Bi$_4$Pd$_3$ \cite{Brass2024}. The locations of the Weyl nodes are obtained by introducing particles with Chern ``charge" at random starting points in the Brillouin zone, which then follow the Berry curvature. Depending on this ``charge", the particles will approach the nodes with either $\mathcal{C}_n=+1$ or $\mathcal{C}_n=-1$. This resembles the motion of charged test particles toward an opposite charge at a fixed position. In this way, it is possible to obtain the accurate locations of the Weyl nodes at much lower computational costs.

\section{Results} 
\label{Sec:Results}
\subsection{DFT+DMFT calculations}
\label{Sec:resultsDFTDMFT}
\begin{figure*}
\centering
\includegraphics[width=0.9\linewidth]{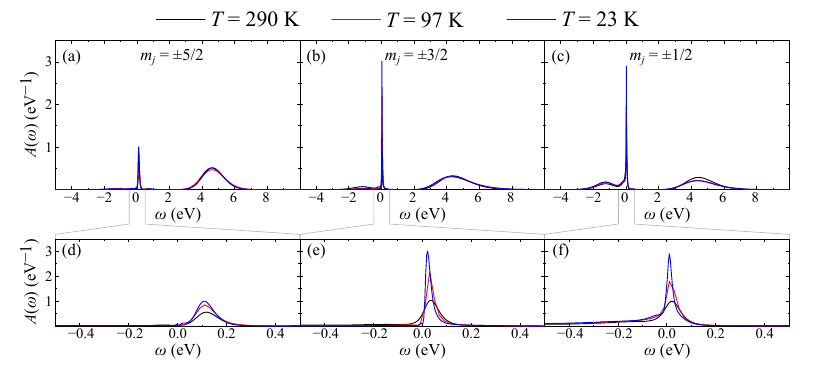}
\caption{DMFT spectral functions $A(\omega)$ for the $3\times 2$ Ce $f_{5/2}$ states at three different temperatures: $T=290\,\mathrm{K}$ (black), $T=97\,\mathrm{K}$ (red), $T=23\,\mathrm{K}$ (blue). Results are resolved by total angular momentum projections: $m_j=\pm5/2$ (a,d), $m_j=\pm3/2$ (b,e) and $m_j=\pm1/2$ (c,f). Graphs (a-c) show the full energy range, while (d-f) are zoomed-in views near the Fermi energy.}
\label{DMFT A(w)}
\end{figure*}

Fig.~\ref{DMFT A(w)} shows the DMFT-computed spectral functions for the Ce $f_{5/2}$ orbitals at three different temperatures: $290\,\mathrm{K}$, $97\,\mathrm{K}$, and $23\,\mathrm{K}$. As the temperature decreases, a spectral peak develops at the Fermi energy and strongly increases in height for the $m_j=\pm3/2$, $\pm1/2$ components [see Figs.~\ref{DMFT A(w)}(e) and \ref{DMFT A(w)}(f)].
This behavior is characteristic of the gradual formation of the Kondo resonance at low temperatures, which in our case is saturated at $23\,$K [the spectral function at 50\,K (not shown) is essentially the same as for 23\,K]. In contrast, the $m_j=\pm5/2$ components exhibit a significantly weaker temperature dependence, showing only minor changes in the studied energy range [Figs.~\ref{DMFT A(w)}(a) and \ref{DMFT A(w)}(d)].

\begin{figure}[tb]
\centering
\includegraphics[width=.99\linewidth]{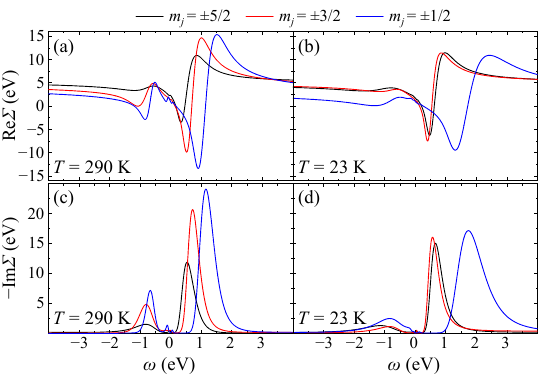}
\caption{Real (a,b) and imaginary (c,d) parts of the DMFT self-energy $\mathit{\Sigma}(\omega)$ at two different temperatures: $290$ $\mathrm{K}$ (a,c), and $23$ $\mathrm{K}$ (b,d). Results are resolved for different $m_j$ of the Ce $4f_{5/2}$ manifold.}
\label{DMFT Sigma(w)}
\end{figure}

Fig.~\ref{DMFT Sigma(w)} presents the real and imaginary parts of the self-energy for different $m_j$ projections of the Ce $j=5/2$ manifold. At room temperature, each $m_j$ component shows different intensities and pole positions. The latter correspond to a (damped) $1/\omega$ behavior of the real part of the self-energy in Fig.~\ref{DMFT Sigma(w)}(a), and a maximum in the imaginary part in Fig.~\ref{DMFT Sigma(w)}(c). These poles split off the Hubbard bands from the quasiparticle band, and are more pronounced for $\omega>0$, where the splitting between the quasiparticle band and the upper Hubbard band is larger (which requires a larger prefactor of the $1/\omega$ pole). The $m_j=\pm1/2$ component exhibits minor glitches near $\omega=0$, which likely arise from numerical artifacts in the analytic continuation. 

At lower temperatures ($23$ $\mathrm{K}$), the pole of $\mathit{\Sigma}(\omega)$ for $m_j=\pm1/2$ is shifted to higher energies [Fig.~\ref{DMFT Sigma(w)}(b)], and the region where $\mathrm{Im}\mathit{\Sigma}(\omega)$ is small at the Fermi energy becomes broader [Fig.~\ref{DMFT Sigma(w)}(d)].
This indicates enhanced quasiparticle lifetimes, concomitant with the development of a sharp Kondo resonance.
The self-energy of the other orbitals ($m_j=\pm5/2,\pm3/2$) becomes nearly degenerate at lower temperatures.

From the DMFT self-energy, we computed $Z$ and $\mathrm{Re}\mathit{\Sigma}(0)$ for each $m_j$ projection of the Ce $4f_{5/2}$ manifold at two temperatures: $290$ $\mathrm{K}$ and $23$ $\mathrm{K}$. For the results, see Table~\ref{ZandS0}. In all cases $Z$ remains similar and approximately $0.1$. On the other hand, $\mathrm{Re}\mathit{\Sigma}(0)$ is nearly temperature independent for the $m_j=\pm1/2$ bands, while it is largest for the $m_j=\pm5/2$ manifold. For $m_j=\pm3/2,\pm5/2$, $\mathrm{Re}\mathit{\Sigma}(0)$ becomes smaller as the temperature is reduced. Also shown in Table~\ref{ZandS0} is the occupation of these different 4$f$ orbitals. With lowering temperature, we see that the $m_j=\pm 1/2$ occupation increases towards half-filling, whereas the other orbitals depopulate. This agrees with previous DFT+DMFT calculations at higher temperatures \cite{Wissgott} and x-ray spectroscopy~\cite{Sundermann2015}.

\begin{table}[tb]
\caption{Calculated quasiparticle parameters and electron filling for the full Ce $f_{5/2}$ manifold from DMFT calculations at two different temperatures: $290$ $\mathrm{K}$ and $23$ $\mathrm{K}$. Chemical potential $\mu$ is for the $90$ orbital basis set.}

\centering
\renewcommand{\arraystretch}{1.3}
\begin{tabular*}{\columnwidth}{@{\extracolsep{\fill}}cccccc}
    \toprule
    $T$ (K) & $m_j$ & $\mu$ (eV) & $Z$ & $\mathrm{Re}\,\mathit{\Sigma}(0)$ (eV) & $\braket{n}$\\
    \midrule
    \multirow{3}{*}{290} 
    & $\pm 1/2$ & \multirow{3}{*}{$-0.080$} & $0.132$ & $0.951$ & $0.569$\\
    & $\pm 3/2$ & & $0.107$ & $1.336$ & $0.293$\\
    & $\pm 5/2$ & & $0.126$ & $2.217$ & $0.096$ \\
    \midrule
    \multirow{3}{*}{23} 
    & $\pm 1/2$ & \multirow{3}{*}{$-0.059$} & $0.112$ & $1.013$ & $0.734$\\
    & $\pm 3/2$ & & $0.104$ & $1.257$ & $0.095$\\
    & $\pm 5/2$ & & $0.123$ & $1.754$ & $0.107$\\
    \bottomrule
\end{tabular*}
\label{ZandS0}
\end{table}

\begin{figure}[tb]
    \centering
    \includegraphics[width=0.85\linewidth]{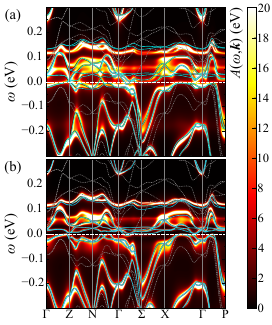}
    \caption{Total DMFT spectral function $A(\omega,\bm{k})$ of all bands at (a) $290$ $\mathrm{K}$ and (b) $23$ $\mathrm{K}$. Overlaid white dashed lines are the initial DFT bands; blue solid lines are the renormalized quasiparticle bands. The latter are in good agreement with the DMFT $A(\omega,\bm{k})$ around the Fermi energy.}
    \label{SpectraFull}
\end{figure}

Fig.~\ref{SpectraFull} shows the $\bm{k}$-resolved spectral function at $290\,\mathrm{K}$ and $23\,\mathrm{K}$.
The original DFT bands are overlaid; they are gapped, representing an insulator. Importantly, they fail to describe the correlated spectral function $A(\omega,\bm{k})$ around the Fermi energy. This is because they lack the Kondo resonance (or quasiparticle peak). In contrast, the overlaid renormalized quasiparticle bands provide a reliable representation of the many-body spectral function at low energies. 

\begin{figure}[tb]
    \centering
    \includegraphics[width=0.85\linewidth,trim=0.15cm 0 0 0]{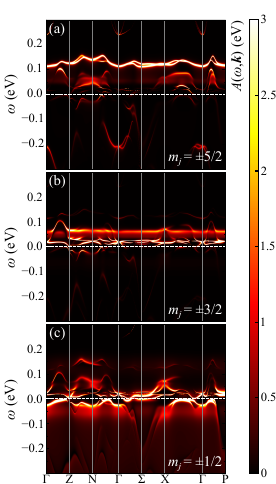}
    \caption{Contribution of the different Ce 4$f_{5/2}$ orbitals to the spectral functions at $23$ $\mathrm{K}$: $m_j=\pm5/2$ (a), $m_j=\pm3/2$ (b) and $m_j=\pm1/2$ (c).}
    \label{ResolvedSpectralFor Beta500}
\end{figure}

At $T=23\,\mathrm{K}$ (Fig.~\ref{SpectraFull}(b)), within the low-energy quasiparticle band, a direct gap of $\sim 5\,\mathrm{meV}$ appears at most wave vectors across the Brillouin zone, in excellent agreement with experiments \cite{Bru06.1}. This gap becomes visible only at low temperatures  where  lifetimes are long enough so that the  resolution of the bands improves. The reason for the gap opening is the hybridization of the Ce 4$f$ states with the conduction bands---the hallmark of a Kondo insulator. However, because the gap is very small and the bands disperse, there is no indirect band gap across the entire Brillouin zone: from Fig.~\ref{SpectraFull}, as well as the $m_j$-resolved spectrum in Fig.~\ref{ResolvedSpectralFor Beta500}, it can be seen that there are bands that cross the Fermi level. 

As we will show in Fig.~\ref{WeylBands} below, electronic correlations lead, in addition, to {\em topological} band crossings of the Kondo insulator gap, with a linear dispersion around the Weyl points. 
These Weyl points lie, however, not on the high-symmetry lines displayed in Fig.~\ref{SpectraFull}.
Note that in Ref.~\cite{Wissgott}, which used Hirsch-Fye QMC as the DMFT impurity solver, such low temperatures could not be reached. The use of continuous-time QMC as well as the improved Weyl node search algorithm  are  essential for the advance presented here.

\subsection{Topological analysis}
\label{Sec:ResultsTopology}
The search for Weyl nodes in \Ce\ at the Fermi level is carried out using the Weyl node search algorithm of~\cite{MartinWeyl}. We focus on the physically most relevant Ce $4f_{5/2}$ orbital with $m_j=(-)1/2$ and used the renormalized quasiparticle Hamiltonian (\ref{renormalized}) with DMFT self-energy at $T=23\,$K as input. Additionally, FPLO-generated primitive unit cell is used, with all $\bm{k}$-points expressed in fractional units of the primitive reciprocal lattice vectors. In total, we identify five inequivalent low-energy Weyl points, which are presented in Table \ref{WeylpointTable}. Some of these are located above the Fermi level, and others below, with the closest node only $0.5$\,meV (6\,K) below the Fermi level.

With the eight symmetry operations of the $D_{2d}$ point group in real space and pure time-reversal in momentum space, the symmetry transformations generate 16 Weyl nodes for every inequivalent Weyl point. 
However, Weyl points I and IV in Table \ref{WeylpointTable} are high symmetry points and thus generate only 8 symmetry-related nodes. All transformation operations are discussed in Appendix~\ref{App:D2d}. Some symmetry-related nodes at the same energy and opposite Chern number are relatively close to each other, see ${\mathbf \Delta k}_{\mathrm min}$ in Table~\ref{WeylpointTable}.

Altogether, this results in a total of $3 \times 16 +2\times 8= 64$ Weyl points, with half carrying Chern number $+1$ and the other half $-1$. Not all of them were detected in the DFT+Gutzwiller band structure \cite{XuHeavyFermions}, which could be due to: (i)  The more advanced DMFT treatment. Note that the Gutzwiller approximation corresponds to a DMFT approximation for the Gutzwiller wave function \footnote{See Chapter 1 of \cite{Juelich2014}}. Given that Xu {\em et al.} \cite{XuHeavyFermions} found Weyl nodes close to high symmetry lines with e.g.\ $k_y\approx 0$ it is (ii) also possible that the improved search algorithm used in the present work helped us identify more Weyl points.

\begin{figure}[h!]
    \centering
    \includegraphics[width=0.85\linewidth]{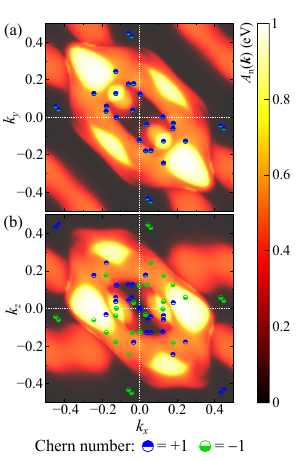}
    \caption{Projection of the Fermi surface spectral function normalized as $A_{\mathrm{n}}(\bm{k})=A(\bm{k},\omega)/\mathrm{max}(A(\bm{k},\omega))$: (a) in the $[k_x,k_y,0]$ plane and (b) in the $[k_x,0,k_z]$ plane. Projected Weyl nodes are marked by blue and green circles, corresponding to the Chern numbers $+1$ and $-1$, respectively.
    }
    \label{Weyl Surf}
\end{figure}

Fig.~\ref{Weyl Surf} displays the Weyl points in a projection onto $k_z=0$ and $k_y=0$ in panels (a) and (b), respectively. Blue and green circles indicate Chern numbers of $+1$ and $-1$; overlaid are the projected DMFT spectral functions computed at the Fermi energy. For both the Weyl points and the spectrum, the projections respect the $D_{2d}$  symmetries around the $k_x=k_y$ and $k_x=-k_y$ planes. The density of the Weyl nodes is larger in the vicinity of the $\Gamma$ point.

\begin{table*}[tb]
  \caption{Fractional coordinates of the inequivalent Weyl nodes in momentum space of a primitive unit cell. Also shown are the total number symmetry-related nodes $N$, their type (I or II), minimal distance $|\bm{\Delta k}_{\mathrm{min}}|$ between symmetry-related opposite Chern charge nodes, energy relative to the Fermi level, geometric mean of the Weyl velocities $(v_xv_yv_z)^{1/3}$, and Ce-$4f_{5/2}$ spectral weight $w_{4f_{5/2}}$. Only one set of nodes with Chern number $+1$ are listed; others can be obtained by symmetry transformations (see Appendix \ref{App:D2d}). Chern numbers are computed by integrating the Berry curvature over a sphere of radius $0.01$, centered at each node. Node types, node energies, and Weyl velocities are obtained from the band structure in the vicinity of the nodes, with the velocities computed from the linear slope of the bands along Cartesian directions. Spectral weight of Ce-$4f_{5/2}$ is extracted from $A(\omega,\bm{k})$ at the corresponding node coordinates.}
\begin{tabular*}{\textwidth}{@{\extracolsep{\fill}}@{\hspace{1em}}c@{\hspace{1em}}|cccc@{\hspace{1em}}|ccc@{\hspace{1em}}}
    \hline
    \text{Node} &\text{Coordinates} & $N$ & \text{Type} & $|\bm{\Delta k}_{\mathrm{min}}|$ (\AA{}$^{-1}$) & \text{Energy (meV)} & $(v_xv_yv_z)^{1/3}$ (km/s) &   $w_{4f_{5/2}}$ (\%) \\
    \hline
    I & $[+0.032,\,-0.179,\,-0.032]$ & $8$  & I & $0.1442$ & $-1.5$ & $51$ &  $67$\\
    II & $[+0.043,\,-0.432,\,+0.448]$ & $16$ & I & $0.3245$ & $22.1$  & $10$ &  $81$\\
    III & $[+0.244,\,-0.127,\,+0.060]$ & $16$ & I & $0.0021$ & $-0.5$  & $7$ &  $77$\\
    IV & $[-0.127,\,+0.127,\,-0.038]$ & $8$  & I & $0.0519$ & $-3.8$  & $17$ &  $81$\\
    V & $[+0.125,\,+0.001,\,-0.003]$ & $16$ & II & $0.0861$ & $12.2$ & $26$ &  $67$ \\
    \hline
\end{tabular*}

    \label{WeylpointTable}
\end{table*}

\begin{figure}[t]
    \centering
    \includegraphics[width=0.85\linewidth,trim=0.1cm 0 0.25cm 0]{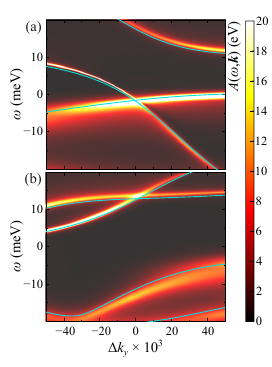}
    \caption{Quasiparticle bands (lines) and DMFT $\bm{k}$-resolved spectral functions (color overlay) in the vicinity of two selected Weyl nodes: (a) type I node at $[0.032;-0.179;-0.032]$ and (b) type II node at $[0.125;0.001;-0.003]$. $\Delta k_y$ is the momentum deviation from the nodes along the $k_y$ direction, while other components are held fixed.} 
    \label{WeylBands}
\end{figure}

In Fig.~\ref{WeylBands}, the dispersion around two exemplary Weyl points of type I and II along the $k_y$ direction is shown at $23\,\mathrm{K}$. For both nodes, the band structure exhibits linearly dispersing crossings close to the Fermi energy, forming Dirac cone-like structures. The Weyl node of type II shown in  Fig.~\ref{WeylBands}(b) is tilted along the (shown) $k_y$ direction. Band crossings in Fig.~\ref{WeylBands} confirm the presence and topological nature of the identified Weyl points. 

From the linear dispersion relations near each Weyl point, we computed the geometric mean of the velocities $(v_xv_yv_z)^{1/3}$ (Table~\ref{WeylpointTable}). These values, in some cases below $10\,\mathrm{km/s}$, are orders of magnitude smaller than those in weakly correlated Weyl semimetals. Such reduced velocities enhance the low-temperature density of states and specific heat, and can significantly amplify nonlinear transport responses.

\section{Conclusion and discussion}
\label{Sec:Conclusion}
We have studied the topology of the low-energy quasiparticle bands of \Ce\ using DFT+DMFT. We find five inequivalent Weyl points or 64 in total close to the Fermi level, bridging the Kondo insulator gap induced by the hybridization between 4$f$ and conduction electrons. This is in agreement with high-field experiments on the related material \CBP, which provided direct evidence for the Weyl nodes being situated within the Kondo insulator gap \cite{Dzs22.1}. Note that unlike in the Weyl-Kondo semimetal model study \cite{Lai18.1,Gre20.1}, where filling constraints pin the Weyl nodes to the Fermi energy, this is not the case here.

In the present study, besides the Kondo temperature, which is of the order of 100\,K in DMFT, additional energy scales are set by: (i) the size of the direct Kondo gap which is about 5\,meV (60\,K), (ii) the energy range by which the bands above and below this direct Kondo gap cross the Fermi energy which is up to about 1\,meV (12\,K), and last but not least (iii) the distance of the various Weyl points from the Fermi energy, the smallest being 0.5\,meV (6\,K). This will lead to a complex temperature dependence, as different phenomena become apparent when crossing the various temperature scales.

Experimentally, \Ce\ was evidenced to be an emergent Weyl-Kondo semimetal, exhibiting a spontaneous nonlinear Hall effect below 1\,K \cite{Kir24.2x}. This is of the same order of magnitude as the smallest separation of a DMFT Weyl node from the Fermi energy (6\,K).
There is also overall agreement in terms of the Weyl velocities, which, both in experiment and DMFT, are reduced by orders of magnitude compared to noninteracting Weyl semimetals.

The $\omega/T$ scaling found in inelastic neutron scattering experiments \cite{Fuh21.1} and motivated by an EDMFT study of a quantum critical Weyl-Kondo semimetal state \cite{Kir24.2x}, indicates Kondo destruction quantum criticality in \Ce\ below a scale of 10\,K . While quantum criticality and Kondo destruction can, in principle, be described by DMFT, we have no indications for such physics in our calculations, which describe a Fermi liquid. However, non-local correlations that are not included in DMFT can compete with the Kondo effect captured in DMFT. This competition can alter the physics if, with decreasing temperature, non-local correlations become strong.

In any case, our theoretical calculations form an \textit{ab initio} basis for understanding the Weyl semimetal nature of CeRu$_4$Sn$_6$. It is a strongly correlated ``failed'' Kondo insulator, with (i) some of the topologically trivial bands crossing the Fermi level and (ii) Weyl nodes bridging the Kondo insulator gap.
Our work highlights the importance of many-body correlations in \Ce\ and heavy-fermion systems in general, and offers a framework for investigating interaction-driven topology, nonlinear transport, and quantum criticality across a wider range of materials.

\begin{acknowledgments}
We thank Markus Wallerberger, Andriy Smolyanyuk, Diana Kirschbaum, and Qimiao Si for very helpful discussions. The support of the Research Unit FOR5249 `QUAST' of the Deutsche Forschungsgemeinschaft [Austrian Science Funds (FWF) project DOI 10.55776/I5868] and the Spezialforschungsbereich (SFB) Q-M\&S FWF project DOI 10.55776/F86 is gratefully acknowledged.
The computational results have been achieved in part using the Austrian Scientific Computing (ASC) infrastructure. For the purpose of open access, the authors have applied a CC BY-NC-SA public copyright license to any Author Accepted Manuscript version arising from this submission.
\end{acknowledgments}

\appendix
\section{$D_{2d}$ symmetry operations in a primitive unit cell}
\label{App:D2d}

To represent the found Weyl nodes, we have chosen a primitive lattice and we express all momentum coordinates in fractional units of the corresponding reciprocal lattice vectors. In order to convert from this lattice to a conventional lattice, one can change coordinate system by applying a transformation matrix.

The unit vectors for a primitive cell of \Ce\ are (in \AA{}):
\[
\begin{array}{rcl}
\mathbf{a}_1 & = & \bigl(\,-6.502,\quad +6.502,\quad +6.502\,\bigr)\,, \\[4pt]
\mathbf{a}_2 & = & \bigl(\,+6.502,\quad -6.502,\quad +6.502\,\bigr)\,, \\[4pt]
\mathbf{a}_3 & = & \bigl(\,+9.214,\quad +9.214,\quad -9.214\,\bigr)\,.
\end{array}
\]

The corresponding reciprocal lattice vectors are (in \AA{}$^{-1}$):
\[
\begin{array}{rcl}
\mathbf{b}_1 & = & \bigl(\, 0.000,\quad 0.483,\quad 0.341 \bigr), \\[4pt]
\mathbf{b}_2 & = & \bigl(\,0.483,\quad 0.000,\quad 0.341 \bigr), \\[4pt]
\mathbf{b}_3 & = & \bigl(\, 0.483,\quad 0.483,\quad 0.000 \bigr).
\end{array}
\]

The transformation matrix $B$ is constructed by placing the reciprocal vectors as columns:
\[
B =
\begin{bmatrix}
b_{1x} & b_{2x} & b_{3x} \\
b_{1y} & b_{2y} & b_{3y} \\
b_{1z} & b_{2z} & b_{3z}
\end{bmatrix} = \begin{bmatrix}
0.000 & 0.483 & 0.483 \\
0.483 & 0.000 & 0.483 \\
0.341 & 0.341 & 0.000
\end{bmatrix}
\]

To map a node from primitive cell coordinates to conventional cell coordinates, the node's coordinate vector must be multiplied by $B$. This transformation yields the coordinates of the Weyl node expressed in the basis of the conventional reciprocal lattice vectors. It is also possible to go from conventional lattice basis to a primitive one by inverting $B$.

$D_{2d}$ symmetry operations also take a different form in the primitive cell. To obtain their representation in reciprocal coordinates consistent with the primitive cell, the conventional symmetry matrices $\alpha_\mathrm{C}$ can be transformed as follows:
\begin{equation}
    \alpha_\mathrm{P}=B^{-1}\alpha_\mathrm{C}B
\end{equation}
This procedure ensures that the symmetry operations act correctly on the reduced $\bm{k}$-space coordinates defined in the primitive basis. Additionally, these transformations are consistent with the changes in the Chern charge associated with the Weyl nodes, ensuring that the symmetry operations correctly map not only the node locations but also their corresponding topological characteristics. All eight $D_{2d}$ symmetry operations in both conventional and primitive cells are summarized in Table~\ref{D2d_symmetry}, together with the effects on the Chern charge of the Weyl nodes. Furthermore, the time-reversal symmetry relates each Weyl node to a partner with the same Chern charge; its momentum can be obtained by changing the sign of all the momentum components. This operation can be combined with the $D_{2d}$ symmetry transformations discussed above, yielding 16 symmetrically related Weyl points in total. However, there is a possibility that Weyl nodes are at high-symmetry points. In this case, only some of the 16 points generated by the symmetry operations are equivalent. This occurs for Weyl points I and IV in Table~\ref{WeylpointTable}.

For example, Weyl node I with $\bm{k}=[+0.032,-0.179,-0.032]$ and Chern number $+1$ (cf.  Table~\ref{WeylpointTable}) has 8 symmetry-related replicas.  Applying, for instance, $C_{2x}$ yields another Weyl node at $\bm{k}=[+0.179,-0.032,-0.179]$ (cf. Table~\ref{D2d_symmetry} for a primitive cell) with the same Chern number $+1$. Applying other transformations such as mirror symmetry ($\sigma_{x=y}$) also yield symmetry-related Weyl nodes with opposite Chern number. Additionally, the combined application of the time-reversal and $C_{2y}$ rotation maps the Weyl node onto itself, $\bm{k}=[+0.032,-0.179,-0.032]$, due to its high-symmetry position.

\begin{table*}[tp]
\centering
\caption{$D_{2d}$ symmetry matrices that transform ${\mathbf k} = (k_x,k_y,k_z)^{\rm Tr}$ on its eight symmetrically related replicas, 
represented in conventional and primitive reciprocal lattice bases, and transformation effect on the Chern charge.}

\begin{tabular*}{\textwidth}{@{\extracolsep{\fill}} l c c c}
\hline
\hline
Symmetry Operation & Conventional Matrix $\alpha$ & Primitive Matrix $B^{-1}\alpha B$ & Chern charge \\
\hline \\[1pt]

Identity ($E$) &

$\begin{bmatrix}
\phantom{-}1 &\phantom{-}0 &\phantom{-}0\\
\phantom{-}0 &\phantom{-}1 &\phantom{-}0\\
\phantom{-}0 &\phantom{-}0 &\phantom{-}1
\end{bmatrix}$ &
$\begin{bmatrix}
\phantom{-}1 &\phantom{-}0 &\phantom{-}0\\
\phantom{-}0 &\phantom{-}1 &\phantom{-}0\\
\phantom{-}0 &\phantom{-}0 &\phantom{-}1
\end{bmatrix}$ & Unchanged \\[20pt]

2-fold rotation around $x$ ($C_{2x}$) &
$\begin{bmatrix}
\phantom{-}1 &\phantom{-}0 &\phantom{-}0\\
\phantom{-}0 &-1 &\phantom{-}0\\
\phantom{-}0 &\phantom{-}0 &-1
\end{bmatrix}$ &
$\begin{bmatrix}
-1 &-1 &-1\\
\phantom{-}0 &\phantom{-}0 &\phantom{-}1\\
\phantom{-}0 &\phantom{-}1 &\phantom{-}0
\end{bmatrix}$ & Unchanged \\[20pt]

2-fold rotation around $y$ ($C_{2y}$) &
$\begin{bmatrix}
-1 &\phantom{-}0 &\phantom{-}0\\
\phantom{-}0 &\phantom{-}1 &\phantom{-}0\\
\phantom{-}0 &\phantom{-}0 &-1
\end{bmatrix}$ &
$\begin{bmatrix}
\phantom{-}0 &\phantom{-}0 &\phantom{-}1\\
-1 &-1 &-1\\
\phantom{-}1 &\phantom{-}0 &\phantom{-}0
\end{bmatrix}$ & Unchanged \\[20pt]

2-fold rotation around $z$ ($C_{2z}$) &
$\begin{bmatrix}
-1 &\phantom{-}0 &\phantom{-}0\\
\phantom{-}0 &-1 &\phantom{-}0\\
\phantom{-}0 &\phantom{-}0 &\phantom{-}1
\end{bmatrix}$ &
$\begin{bmatrix}
\phantom{-}0 &\phantom{-}1 &\phantom{-}0\\
\phantom{-}1 &\phantom{-}0 &\phantom{-}0\\
-1 &-1 &-1
\end{bmatrix}$ & Unchanged \\[20pt]

Mirror in $x=y$ plane ($\sigma_{x=y}$) &
$\begin{bmatrix}
\phantom{-}0 &\phantom{-}1 &\phantom{-}0\\
\phantom{-}1 &\phantom{-}0 &\phantom{-}0\\
\phantom{-}0 &\phantom{-}0 &\phantom{-}1
\end{bmatrix}$ &
$\begin{bmatrix}
\phantom{-}0 &\phantom{-}1 &\phantom{-}0\\
\phantom{-}1 &\phantom{-}0 &\phantom{-}0\\
\phantom{-}0 &\phantom{-}0 &\phantom{-}1
\end{bmatrix}$ & $\mathcal{C}\to\,-\mathcal{C}$ \\[20pt]

Mirror in $x=-y$ plane ($\sigma_{x=-y}$) &
$\begin{bmatrix}
\phantom{-}0 &-1 &\phantom{-}0\\
-1 &\phantom{-}0 &\phantom{-}0\\
\phantom{-}0 &\phantom{-}0 &\phantom{-}1
\end{bmatrix}$ &
$\begin{bmatrix}
\phantom{-}1 &\phantom{-}0 &\phantom{-}0\\
\phantom{-}0 &\phantom{-}1 &\phantom{-}0\\
-1 &-1 &-1
\end{bmatrix}$ & $\mathcal{C}\to\,-\mathcal{C}$ \\[20pt]

4-fold rotoinversion around $z$ ($S^+_{4z}$) &
$\begin{bmatrix}
\phantom{-}0 &\phantom{-}1 &\phantom{-}0\\
-1 &\phantom{-}0 &\phantom{-}0\\
\phantom{-}0 &\phantom{-}0 &-1
\end{bmatrix}$ &
$\begin{bmatrix}
-1 &-1 &-1\\
\phantom{-}0 &\phantom{-}0 &\phantom{-}1\\
\phantom{-}1 &\phantom{-}0 &\phantom{-}0
\end{bmatrix}$ & $\mathcal{C}\to\,-\mathcal{C}$ \\[20pt]

4-fold rotoinversion around $z$ ($S^-_{4z}$) &
$\begin{bmatrix}
\phantom{-}0 &-1 &\phantom{-}0\\
\phantom{-}1 &\phantom{-}0 &\phantom{-}0\\
\phantom{-}0 &\phantom{-}0 &-1
\end{bmatrix}$ &
$\begin{bmatrix}
\phantom{-}0 &\phantom{-}0 &\phantom{-}1\\
-1 &-1 &-1\\
\phantom{-}0 &\phantom{-}1 &\phantom{-}0
\end{bmatrix}$ & $\mathcal{C}\to\,-\mathcal{C}$ \\[20pt]

\end{tabular*}
\label{D2d_symmetry}
\end{table*}

\clearpage


%

\end{document}